\documentstyle[12pt]{article}
\input tcilatex
\begin{document}

\begin{center}
{\bf Three-dimensional interior solutions from Jordan-Brans-Dicke Theory of
Gravity.}

\vspace*{1.5cm} S.M.KOZYREV

e-mail: Kozyrev@e-centr.ru
\end{center}

\vspace*{1.5cm}

\begin{center}
${\bf Abstract}$
\end{center}

{\rm We study spherically symmetric solutions to the Jordan-Brans-Dicke
field equations under the assumption that the space-time may possess an
arbitrary number of spatial dimensions. Assuming a perfect fluid with the
equation of state p = }$\varepsilon \rho ${\rm , we show that there are
static interior nontrivial solutions in three dimensional Jordan-Brans-Dicke
gravity theory.}

\section{Introduction}

The most studied and hence the best known alternative of classical
Einstein's gravity is the scalar-tensor Jordan-Brans-Dicke (JBD) theory.\cite
{1}\cite{2}.The essential feature of JBD theory is the presence of a
massless scalar field to describe gravitation together with the metric.
Scalar-tensor theories contain arbitrary functions of the scalar field that
determine the scalar potential as a dynamical variable, the analog of
gravitational permittivity is allowed to vary with space and time which
defined using Newton's gravitational constant as G = 1/$\phi $.

The investigate of a JBD field equations in dimensions lower than 3+1 is
interesting because it may allow to study phenomena characteristic of
gravity, which have 3+1 dimensional analogues in a simplified context \cite
{3}. In 2+1 dimensions, the Riemann- Christoffel tensor is uniquely
determined by the Ricci tensor, which vanishes outside the sources. Hence,
spacetime of Einstein General Relativity is flat in regions devoid of
matter, and test particles do not feel any gravitational field. However, in
2+1 dimensions JBD field equations reproduces Newtonian gravity when the low
energy regimen is consistently analysed \cite{4}.

In general we expect that the task of finding solutions of field equations
in 3+1 dimensions to be much more involved than in 2+1. Therefore it will be
useful to find a exact interior solutions to the JBD equations from 2+1
dimensions. In this paper we will show that for a static perfect fluid with
the proper energy density proportional to the proper pressure this can be
done.

\section{Scalar-tensor theories in D--dimensions}

Scalar-tensor theories are described by the following action in the Jordan
frame in $D$-dimensional space-time is:

\begin{eqnarray}
S={\frac 1{16\pi }}\int d^Dx\sqrt{-{g}}\left( {\phi R}-\omega (\phi ){g}%
^{\mu \nu }\partial _\mu \phi \partial _\nu \phi -\lambda (\phi )\right)
+S_m.  \label{2.1}
\end{eqnarray}
Here, ${R}$ is the Ricci scalar curvature with respect to the space-time
metric ${g}_{\mu \nu }$. We use units in wich gravitational constant ${\cal G%
}$=1 and speed of light {\it c}=1. The dynamics of the scalar field $\phi $
depends on the functions $\omega (\phi )$ and $\lambda (\phi )$. It should
be mentioned that the different choices of such function give different
scalar tensor theories. We restrict our discussion to the JBD theory which
characterized by the functions $\lambda (\phi )=0$ and $\omega (\phi
)=\omega /\phi $, where $\omega $ is a constant.

The action principle with suitable boundary conditions gives rise to the
D-dimensional field equations:

\begin{eqnarray}
G_{\mu \nu }=-{\frac{8\pi }\phi }T_{\mu \nu }-{\frac \omega {\phi ^2}}(\phi
_{,\mu }\phi _{,\nu }-{\frac 12}g_{\mu \nu }\phi _{,\rho }\phi ^{,\rho })-{%
\frac 1\phi }(\phi _{,\mu ;\nu }-g_{\mu \nu }\phi _{;\rho }^{;\rho }),
\label{2.2}
\end{eqnarray}

\begin{eqnarray}
\phi _{;\rho }^{;\rho }={\frac{{\it k}}{2(\omega +1)}T},  \label{2.3}
\end{eqnarray}
where semi-colon denotes the covariant derivative with respect to the metric
g$_{\mu \nu }$ and T$_{\mu \nu }$ is the usual energy momentum tensor which
obeys the conservation equation T$_{\mu \nu ;\rho }g^{\nu \rho }$ = 0, and $%
k $ is a function of $\omega $ and dimension D:

\begin{eqnarray}
k=\frac{1+\omega }{\left( D-1\right) +\omega \left( D-2\right) }.
\label{2.4}
\end{eqnarray}

The progress in the understanding of scalar-tensor theories of gravity is
closely connected with finding and investigation of exact solutions. We
assume that space-time is static and the configuration is sphericaly
symmetric. In this case futher simplification is possible; the space-time
metric can be put in the curvature coordinates as: 
\begin{eqnarray}
ds^2={e}^{2\beta }dt^2-e^{2\alpha }dr^2-r^2d\Omega _{\left( D-2\right) }^2,
\label{2.5}
\end{eqnarray}

where $d\Omega _{\left( D-2\right) }^2$ is the line element on a unit D-2
sphere:

\begin{eqnarray}
d\Omega _{\left( D-2\right) }^2=\left[ d\theta _{\left( 0\right) }^2+%
\stackrel{D-3}{\stackunder{n=1}{\sum }}d\theta _{\left( n\right) }^2\left( 
\stackrel{n}{\stackunder{m=1}{\prod }}\sin ^2\theta _{\left( m-1\right)
}\right) \right] .  \label{2.6}
\end{eqnarray}

In the case of JBD theory it allow us to find exact solutions for the field
produced by a static and isotropic source in regions devoid of sources in D
= 3. In a suitable reference system, the metric can be written in the
standard form:

\begin{eqnarray}
ds^2=e^{2\beta }dt^2-e^{2\alpha }dr^2-r^2d\varphi ^2,  \label{2.7}
\end{eqnarray}

where $\alpha $ and $\beta $ are functions of $r$ alone. In vacuum (T$_{\mu
\nu }$ = 0) Eqs. (\ref{2.2}), (\ref{2.3}) have the solution g$_{\mu \nu
}=\eta _{\mu \nu }$ and $\phi $ = 1, where $\eta _{\mu \nu }$ is the
Minkowski metric tensor. On the other hand in vacuum Eqs. (\ref{2.2}), (\ref
{2.3}) yield

\begin{eqnarray}
\alpha =\alpha _0+C_\alpha \ln \left( \frac r{r_0}\right) ,  \label{2.8}
\end{eqnarray}
\begin{eqnarray}
\beta =\beta _0+C_\beta \ln \left( \frac r{r_0}\right) ,  \label{2.9}
\end{eqnarray}

\begin{eqnarray}
\phi =\phi _0\left( \frac r{r_0}\right) ^{C_\phi }.  \label{2.10}
\end{eqnarray}
Hawking \cite{5} has pointed out that the stationary space containing a
black hole is a solution of the JBD field equations if and only if it is a
solution of the Einstein field equations. Thus, stationary black hole
solutions in the JBD theory are the same as stationary black hole solutions
in the Einstein theory. However, JBD theory can be thought of as a minimal
extension of general relativity designed to properly accommodate both Mach's
principle \cite{6} and Dirac's large number hypothesis \cite{6}. The point
is that the field equations admit curved vacuum solutions and also solutions
that are asymptotically flat, at the same time such solutions as vacuum, a
single matter particle etc. are anti-machian and, hence, has no meaning in
the absence of matter.

\section{Interior solutions.}

Finding exact solutions of scalar-tensor theories equations in the presence
of a matter is a difficult task due to their complexity in the general case.
Bruckman and Kazes \cite{7} derive the relation between the scalar field $%
\phi $ and g$_{00}$:

\begin{eqnarray}
\phi ={e}^{k\beta },  \label{2.11}
\end{eqnarray}

where {\it k} arbitrary constant. They use static spacetime and the energy
-momentum tensor is that of a perfect fluid with equation of state p ={\rm \ 
}$\varepsilon \rho $. Using this assumption an exact solution of the field
equation in 3+1 dimensions is found corresponding to a density distribution
that is infinite at the origin \cite{7}. Moreover using the relation (\ref
{2.11}) we have shown that the existence of a solution of Eqs. (\ref{2.2}), (%
\ref{2.3}) in 3+1 dimensions with equation of state $\rho $ = 0 and p$\neq $%
0 \cite{8}. However, this standard tenet about the relation between $\phi $
and g$_{00}$ can be false, this relation in general case is more complexity 
\cite{8}.

The equations (\ref{2.2}), (\ref{2.3}) are particularly simple with the
choice space in 2+1 dimensions. In this case we get the field equations of
JBD theory produced by a static and isotropic source in regions devoid of
sources in D = 3. The above metric (\ref{2.7}) yields the following field
equations for (\ref{2.2}), (\ref{2.3}):

\begin{eqnarray}
-\frac{\phi ^{\prime }}r+\alpha ^{\prime }\phi ^{\prime }-\beta ^{\prime
}\phi ^{\prime }-\phi ^{\prime \prime }=\frac{4\pi e^{2\alpha }\left(
2p-\rho \right) }{2+\omega },  \label{2.12}
\end{eqnarray}

\begin{eqnarray}
\frac{\alpha ^{\prime }}r+\alpha ^{\prime }\beta ^{\prime }-\beta ^{\prime
2}+\frac{\alpha ^{\prime }\phi ^{\prime }}\phi -\frac{\omega \phi ^{\prime 2}%
}{\phi ^2}-\beta ^{\prime \prime }-\frac{\phi ^{\prime \prime }}\phi ={\frac{%
8\pi e^{2\alpha }\left( \omega p-\left( 1+\omega \right) \rho \right) }{%
\left( 2+\omega \right) \phi }},  \label{2.13}
\end{eqnarray}

\begin{eqnarray}
r\left( \alpha ^{\prime }-\beta ^{\prime }-\frac{\phi ^{\prime }}\phi
\right) ={\frac{8\pi e^{2\alpha }r^2\left( \omega p-\left( 1+\omega \right)
\rho \right) }{\left( 2+\omega \right) \phi }},  \label{2.14}
\end{eqnarray}
\begin{eqnarray}
\frac{\beta ^{\prime }}r-\alpha ^{\prime }\beta ^{\prime }+\beta ^{\prime 2}+%
\frac{\beta ^{\prime }\phi ^{\prime }}\phi +\beta ^{\prime \prime }=-{\frac{%
8\pi e^{2\alpha }\left( 2\left( 1+\omega \right) p+\rho \right) }{\left(
2+\omega \right) \phi }},  \label{2.15}
\end{eqnarray}
where now the primes stand for derivation respect to {\it r}. Let us start
with the problem of finding out the space-time and the scalar field
generated by a static configuration with the choice of the equation of state
in the form:

\begin{eqnarray}
p=\frac{1+\omega }\omega \rho .  \label{2.16}
\end{eqnarray}

Following the previous reasoning all we have to do is to solve equations (%
\ref{2.12})-(\ref{2.15}), which taking into account (\ref{2.16}), reduces to:

\begin{eqnarray}
{\alpha }=C1+\beta +\ln \phi ,  \label{2.17}
\end{eqnarray}
\begin{eqnarray}
\beta =C3+C2\ln r+\left( 2+4\omega \right) \ln \phi ,  \label{2.18}
\end{eqnarray}
\begin{eqnarray}
\phi =r^{\frac{-3-C2-4\omega +k1}{4+7\omega }}\left( C4+r^{\frac{2k1}{%
3+4\omega }}\right) ^{-\frac{3+4\omega }{4+7\omega }}C5,  \label{3.19}
\end{eqnarray}
\begin{eqnarray}
\rho =k2\left( C4+r^{\frac{2k1}{3+4\omega }}\right) ^{\frac{\left( 1+2\omega
\right) \left( 7+16\omega \right) }{\left( 4+7\omega \right) }}r^{k3},
\label{3.20}
\end{eqnarray}
where $C1,C2,C3,C4,C5$ is an arbitrary constants and {\it :}

\begin{eqnarray*}
k1 &=&\sqrt{C2^2-2C2\left( 1+3\omega \right) +\left( 3+4\omega \right) ^2},
\\
k2 &=&\frac{C4\cdot e^{-2\left( C1+C3\right) }C5^{-5-8\omega }\omega \cdot
k1^2}{\pi \left( 3+4\omega \right) \left( 4+7\omega \right) }, \\
k3 &=&\frac{\left( 1+2\omega \right) \left( 3C2\left( 3+4\omega \right)
-\left( 3-k1+4\omega \right) \left( 7+16\omega \right) \right) }{\left(
3+4\omega \right) \left( 4+7\omega \right) }.
\end{eqnarray*}

As a second example let us consider the equation of state which describes
the space-time generated by a matter with $\rho =0$ and p $\neq 0$. In this
case, there is not analogous solutions in classical Einstein's gravity. The
exact solution we can find using the value of $\omega =0.$ Then, from (\ref
{2.2}), (\ref{2.3}) it follows that the sought-for line element, which
describes the space-time generated in JBD theory, is given by

\begin{eqnarray}
{\alpha }=C1+\beta +\ln \phi ,  \label{2.20}
\end{eqnarray}

\begin{eqnarray}
\beta =C3+C2\ln r+2\ln \phi ,  \label{2.21}
\end{eqnarray}

\begin{eqnarray}
\phi =\frac{r^{\frac 14\left( -3-C2+k1\right) }C5}{\left( r^{\frac
23k1}+C4\right) ^{\frac 34}},  \label{2.22}
\end{eqnarray}

\begin{eqnarray}
p=k2\left( C4+r^{\frac{2k1}3}\right) ^{\frac 74}r^{k3},  \label{2.23}
\end{eqnarray}

where $C1,C2,C3,C4,C5$ is an arbitrary constants and {\it :}

\begin{eqnarray*}
k1 &=&\sqrt{9+C2\left( C2-2\right) }, \\
k2 &=&\frac{C4e^{-2\left( C1+C3\right) }k1^2}{12\pi C5^5}, \\
k3 &=&\frac{21-9C2-7k1}{12}.
\end{eqnarray*}

\section{Conclusions}

We have considered spherically symmetric interior solutions to the JBD field
equations with a 2+1 number of spatial dimensions. The property of this
field has been illustrated by computing the metric for the special equation
of state (\ref{2.16}). A reasonably method has been presented which allows
one to solve, at least in this case, these equations. Moreover, using this
method we find solution with $\rho =0$ and p $\neq 0$ that has not analogous
in classical Einstein's gravity. Finally, it is worth mentioning that the,
in contrast with results of Bruckman and Kazes \cite{7}, relation between
the scalar field $\phi $ and g$_{00}$ is more complexity then $\phi ={e}%
^{k\beta }$.\\
\[
\\{\bf Acknowledgements}\\
\]
\\I am grateful to Stoytcho S. Yazadjiev for stimulating discussion.

\end{document}